# A secure approach for embedding message text on an elliptic curve defined over prime fields, and building 'EC-RSA-ELGamal' Cryptographic System

Ahmad Steef
Department of Mathematics,
Al-Baath University, Homs, Syria
e-mail: msma776@gmail.com

M. N. Shamma
Arab International University(AIU), Syria &
Basic Sciences Department, the Mechanical and
Electrical Engineering, Damascus University, Syria
e-mail: Shamman01@yahoo.com

A. Alkhatib
Department of mathematics,
Al-Baath University, Homs, Syria
e-mail: abdulbaset1962@yahoo.com

*Abstract*— **This paper presents a new probabilistic approach to embedding message text on an elliptic curve, by using the concept of the RSA Algorithm and its security, and such approach allows us discovering the message from the point, only according to the security of the RSA Algorithm. By mixing between the concept of this approach and the concept of EC-ELGamal Cryptographic System, we have built a cryptographic System and we named it 'EC-RSA-ELGamal'**

*Keywords RSA Cryptographic System; EC-ELGamal Cryptographic System; Elliptic Curve over finite field; Elliptic Curve Cryptography(ECC); Embedding message text on an elliptic curve.*

## 1. Introduction

The Cryptography is the most important science have being used to secure data while transmitting in networks, so, and because of the increasing of progress of technology, the researchers are working to offer the best techniques and scientific approaches for achieving the most secure steps, and all those depending on the concepts that related of many topics in advanced mathematics.

Cryptography, basically, is divided into two basic types; symmetric cryptography(the encryption and decryption operations use one secure key), and Asymmetric cryptography(the encryption and decryption operations use two keys; one called private and the other is public). The cryptographic system which uses the symmetric approach is called Symmetric Cryptographic System, and the other is called Asymmetric Cryptographic System(or public Cryptographic Systems). The Asymmetric Cryptographic Systems are considered more secure than symmetric. The security of those systems depends on some open problems in both mathematics and computer science, and for example; the security of RSA Cryptographic System depends on: Integer Factorization Problem(IFP), and the security of ElGamal

Cryptographic System depends on Discrete Logarithm Problem(DLP) [1,2].

There are special kinds of algebraic curves called Elliptic Curves. They have been using with many of the famous Asymmetric Cryptographic algorithms, and the first time was in 1985, by Neal Koblitz and Victor Miller[7,6]. That using generates a special kind of cryptographic systems which called elliptic curve cryptographic systems(ECC). The security of those systems depends basically, on Elliptic Curve Discrete Logarithm Problem(ECDLP), and this problem is more difficult than others like IFP and DLP, because so far there has not been found any 'subexponential-time Algorithm' for solving ECDLP, and there are just 'exponential-time Algorithm', while there are 'subexponential-time Algorithms' for solving problems like IFP and DLP. And thus the 'ECC' systems are the most secure ones in these days. For that reason, many researchers work on offering several approaches by using the concepts of elliptic curves with Asymmetric algorithms to reaching to the highest degrees of security and protection of information.

There are various kinds of 'ECC Systems', and for instance we mention the most famous and applicable ones; EC-Diffie Helman, and EC-ElGamal Cryptographic Systems.

To do encryption and decryption message texts by ECC systems, we need using some approaches to embedding the message on an elliptic curve firstly, and then applying the principles which related of encrypting and decrypting operations by those systems.

In this paper we presented a new and secure approach for embedding message text on an elliptic curve (i.e: mapping message text onto a point on an elliptic curve) which defined over prime fields, and then we built some cryptographic system. This system is a mix between the concept of EC-ElGamal system and the concept of RSA system, and thus this system depends on the two problems IFP and ECDLP at the same time.

## 2 . Preliminaries and background

### 2.1 Elliptic Curve over prime fields

According to [3,4], EC over prime field $F_p = Z_p$ and the group law are defined as the following:

EC over prime field $F_p$ is a special kind of algebraic curves given by the following equation (1):

$$y^2 = x^3 + ax + b \pmod{p} \qquad (1)$$

with the condition $4a^3 + 27b^2 \not\equiv 0 \pmod{p}$. This kind of curves consists of all points which satisfy that equation (1) with some point called 'point at infinity'; '$O$'. We can define EC over prime field as form:

$$E(F_p) = \{(x, y) \in F_p \times F_p; \ y^2 = x^3 + ax + b \pmod{p};$$

$$a, b \in F_p \text{ and } 4a^3 + 27b^2 \not\equiv 0 \pmod{p}\} \cup \{O\}$$

The set $E(F_p)$ with a special binary operation '$+$' forms an algebraic group and the identity is '$O$'. The operation '$+$' satisfies the properties ' group law ':

• $O + P = P, \ \forall p \in E(F_p)$

• The inverse of the point $P(x, y) \in E(F_p)$ is:

$-P = (x, -y), and \ P + (-P) = O$

• Point addition:

Let $P(x_1, y_1), Q(x_2, y_2) \in E(F_p); \ P \neq \pm Q$ then;

$P + Q = (x_3, y_3);$

$$x_3 = \left(\frac{y_2 - y_1}{x_2 - x_1}\right)^2 - x_1 - x_2, \qquad (2)$$

$$y_3 = \left(\frac{y_2 - y_1}{x_2 - x_1}\right)(x_1 - x_3) - y_1, \qquad (3)$$

• Point doubling:

Let $P(x_1, y_1) \in E(F_p); \ P \neq -P$ then;

$P + P = 2P = (x_3, y_3);$

$$x_3 = \left(\frac{3x_1^2 + a}{2y_1}\right)^2 - 2x_1, \qquad (4)$$

$$y_3 = \left(\frac{3x_1^2 + a}{2y_1}\right)(x_1 - x_3) - y_1, \qquad (5)$$

**Note 1** all calculations above are computed module $p$.

### 2.2 EC-ELGamal Cryptographic system

Depending on [1], we can illustrate the technique of this system and how using it for encryption and decryption message text between two sides; 'Bob' is the sender of message '$M$', and the receiver 'Alice', as the following steps:

• Step 1: In the first, Alice and Bob should agree to the elliptic curve $E$ over $F_p$, and some point

$P \in E(F_p)$ which generates some large subgroup from $E(F_p)$ with order $n$ and so, the information $(E, P, p)$ is public for Alice and Bob.

• Step 2: Alice selects number $a$ randomly; $1 < a < n$ and computes $Q = aP \pmod{p}$, and sends this to Bob.

• Step 3: Bob selects number $b$ randomly; $1 < b < n$, and computes $bP \pmod{p}$, (M+bQ) $\pmod{p}$, and sends the cipher $C = (bP, M + bQ) \pmod{p}$ to Alice.

• Step 4: Alice decrypts the message by applying the following: $(M + bQ - a(bP)) \pmod{p} \equiv M$

The Figure (1) [1] illustrates all steps above.

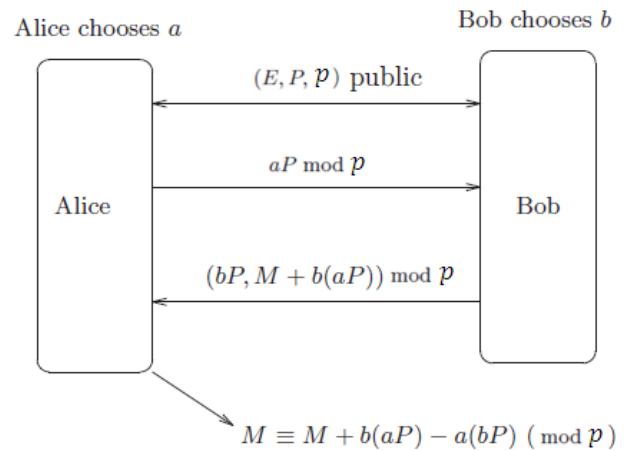

**Figure 1** Description of 'EC-ALGamal Cryptographic System'

**Note 2** Any message text can be represented as number or sets of numbers according to the encoding system which we use it, so from now when we mention the word 'message' , we mean some number.

### 3.2 RSA Cryptographic System [1,2]

RSA Cryptographic System is the most common applications of Public-Key Cryptography(Asymmetric systems). It was published by Rivest, Shamir and Adelman 1978. It uses two distinct keys, public-key which possible to be known to everyone and the other is private-key which is secured and not allowed to exchange between the sender and receiver. The security of this system depends on some open problem in both mathematics and computer science which is

'Integer Factorization Problem(IFP)'. This system described as following:

1. Choose two distinct large random prime numbers $p$ and $q$ .

2. Compute $n = p.q$

3. Compute Euler's function of $n$; $\phi(n)=(p-1).(q-1)$ .

4. Choose an integer $e$ such that:

$1 < e < \phi(n)$, and $\gcd(e, \phi(n)) = 1$

5. Compute $d$ such that: $e.d = 1(\text{mod } \phi(n))$ .

$(e, n)$ is the public – key and $d$ is the private –key.

To encrypt Message ' $M$ ', and get cipher ' $C$ ' by this system we use equation:

$$C = M^e (\text{mod n}) \qquad (6)$$

To do decryption, and coming back to message ' $M$ ' we use equation:

$$C^d (\text{mod n}) = M , \qquad (7)$$

The figure (2) [1] illustrates the RSA System and how it is used between two sides; 'Alice and Bob' to encrypt and decrypt the message ' $M$ '.

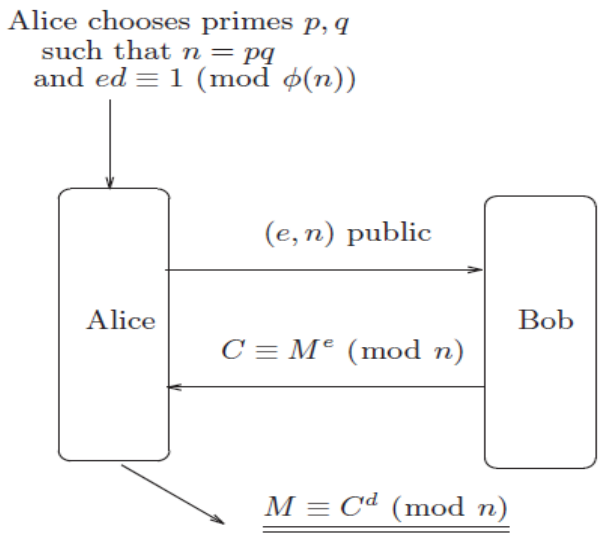

**Figure 2** Description of 'RSA Cryptographic System'

### 3. Proposed approach for mapping message text onto a point on an elliptic curve defined over prime field

The mapping (or embedding) message (some number) onto a point on an elliptic curve is an important approach to do encryption and decryption message texts in every ECC system.

The references [1,3], presented the most practical approach known for mapping message text onto a point on an elliptic curves which defined over finite fields, and this approach attributed to Koblitz as mentioned in [3,5], and according to [1,3], we shall illustrate this approach as following:

Let elliptic curve; $E ; y^2 = x^3 + ax + b(\text{mod p})$ , and suppose $M, K$ are positive integers such that $(M + 1).K < p$ , so, we want to mapping the number $M$ onto some point on $E$ , and to do that just we apply the following:

- We compute a set of values of $x$ ; $x = \{KM + j ; \ j=1,2,3....k-1\}$ until finding some value for $x$ satisfies the equation of $E$ above(i.e: $x^3 + ax + b$ becomes quadratic residue modulo $p$ ), and then we stop and getting the point $(x, y) \in E(F_p); y^2 = x^3 + ax + b(\text{mod p})$ , and thus the number $M$ is being mapped to the point $(x, y)$ .

- To coming back to the number $M$ from the point $(x, y)$ just we can note that $M = \left\lfloor \dfrac{x}{K} \right\rfloor$ ; $\left\lfloor \dfrac{x}{K} \right\rfloor$ is the biggest integer number less than or equals to $\dfrac{x}{K}$

**Theorem 1** If $p$ be an odd prime, then there are exactly $\dfrac{p-1}{2}$ quadratic residues and exactly $\dfrac{p-1}{2}$ quadratic nonresidues modulo $p$ [1].

According to theorem (1), we deduce that the probability of $x^3 + ax + b$ is to be a quadratic nonresidue modulo $p$ for all $K$ values of $x$ is less than or equals to $\dfrac{1}{2^K}$ .

For mapping message text onto a point on elliptic curve, it was easy to coming back from the point to message text(it is easy to discovering the message text from the point) like what we showed above, and for that reason, we shall present a new approach and see how it does not allow 'to whom are not have permission' from discovering the message text from the point.

Our approach depends on using RSA System with elliptic curve defined over prime fields and searching 'in a probabilistic way' for some point and embedding the message text to that point. Our Proposed Approach is illustrated as following:

Let elliptic curve; $E ; y^2 = x^3 + ax + b(\text{mod p})$ , and suppose the number $M$ we want to mapping it onto some point on $E$ , and to do that we apply the following steps:

- Step 1: select two distinct prime numbers $q$ and $r$ such that $M < n = q.r < p$ , and then move to step 2.

- Step 2: find $e$ $1 < e < \phi(n)$, $\gcd(e, \phi(n))=1$ and move to step 3.

- Step 3: compute the value $x = M^e (\text{mod n})$ , then move to step 4.

- Step 4: if $x$ satisfies the equation of $E$ above(i.e: $x^3 + ax + b$ is a quadratic residue modulo $p$ ), then move to step 5, else move to step 2.
- Step 5: stop and the number $M$ is being mapped to the point $(x, y) \in E(F_p); y^2 = x^3 + ax + b(\text{mod p})$ .

To coming back from the point $(x, y)$ to the number $M$ , we have to know the number $d$ ; $e.d = 1(\text{mod } \phi(n))$, then we apply the following:

$$M = x^d (\text{mod n})$$

**Note 3** In our approach above, the number $d$ is not allowed to any one does not have permission. Just whom have permission can discover the number $d$ from the point $(x, y)$ . And discovering it is dependent on the security of RSA algorithm which depends basically on the problem 'IFP'.

**Theorem 2** if $p$ is odd and prime number, then the number $a$ is a quadratic residue modulo $p$ if and only if
$$a^{\frac{p-1}{2}} = 1(\text{mod p})$$ [1].

This theorem (2), helps us to investigate either the number $a$ is a quadratic residue modulo $p$ or not, and so, we use it to see if $x^3 + ax + b$ is a quadratic residue modulo $p$ or not and thus if $x$ presents a $x$ - coordinate for some point on EC or not.

### 3.1 Numerical Example 1

Let the elliptic curve;
$E$ ; $y^2 = x^3 + 71x + 602(\text{mod } 1009)$ [1], and suppose the number $M = 439$ we want to mapping it onto some point on $E$ , and to do that we apply the following steps:
- Step 1: Let $q = 23$, r=43 $\Rightarrow$ n=989, $\phi(\text{n})$=924
- Step 2:select $1 < e < \phi(n)$, and $\gcd(e, \phi(n)) = 1$
  Suppose $e = 5 \Rightarrow$
- Step 3: $x = M^e (\text{mod n})$=354 $\Rightarrow$
- Step 4: $x^3 + 71x + 602$ is a quadratic residue modulo p= 1009
- Step 5: stop and the number $M = 439$ is being mapped to the point $(x, y) = (354, 88) \in E(F_{1009})$;
  $$y^2 = x^3 + ax + b(\text{mod } 1009)$$

To coming back from the point $(354, 88)$ to the number $M = 439$ , we have to know the number $d$ ; $e.d = 1(\text{mod } \phi(n))$ , and then we can apply the following:
$$x^d (\text{mod n})=M$$
By Extended Euclidian Algorithm for instance we have $d = 185 \Rightarrow$

$$(354)^{185} (\text{mod } 989)=439=\text{M}. \quad \checkmark$$

### 4. Building a new cryptographic system 'EC-RSA-ELGamal'

Depending on the ideas showed in paragraphs (3.2) and (3) above, we shall present a cryptographic system and how using it to do encryption and decryption.

Since, this system basically, depends on the concepts of RSA and EC-ELGamal Systems, so we named it 'EC-RSA-ELGamal' system.

Let elliptic curve; $E$ ; $y^2 = x^3 + ax + b(\text{mod p})$ , and suppose $(x, y) \in E(F_p)$ which generates some large subgroup from $E(F_p)$ and its order is $n$ . Suppose the number $M$ we want to encrypt and decrypt it by our cryptographic system. To do that, we apply the following steps:
- Step 1: select two distinct prime numbers $q$ and $r$ such that $M < n = q.r < p$ , and then move to step 2
- Step 2: find $e$ $1 < e < \phi(n)$, $\gcd(e, \phi(\text{n}))=1$ and move to step 3.
- Step 3: compute the value $x = M^e (\text{mod n})$, then move to step 4.
- Step 4: if $x$ satisfies the equation of $E$ above(i.e: $x^3 + ax + b$ is a quadratic residue modulo $p$ ), then move to step 5, else move to step 2.
- Step 5: The number $M$ is being mapped to the point $A_0(x_0, y_0) \in E(F_p); y_0^2 = x_0^3 + ax_0 + b(\text{mod p})$ , and move to step 6.
- Step 6: Compute the point $Q(x_Q, y_Q)$ as form:

$$Q(x_Q, y_Q) = A_0 + a_1 b_1 P \ (\text{mod p}); \ 1<a_1, b_1 < s \qquad (8)$$

and point $Q$ includes the cipher.

To do decryption, and coming back from the point $Q$ to the number $M$ , we apply the following steps:
- Step 1: Compute the point $A_0(x_0, y_0)$ by using the equations (2,3,4,5) as following:

$$(x_Q, y_Q) + (-a_1 b_1 P) \ (\text{mod p}) = A_0 = (x_0, y_0) \qquad (9)$$

- Step 2: Compute the message $M$ by applying the equation following:
  $$x_0^d (\text{mod n})=\text{M}; \ e.d = 1(\text{mod } \phi(n))$$

**Note 4** We can note that EC-RSA-ELGamal Cryptographic System mentioned above offers alternative factor for security if the point $(x, y)$ has been known by someone does not have permission, especially according to the increasing in progress of technology, thus if the point $(x, y)$ is caught 'by some way' by some eavesdropper, there is additional factor for security depending on the security of RSA algorithm .

### 4.1 Numerical Example 2

Let elliptic curve;

$E$; $y^2 = x^3 + 71x + 602 \pmod{1009}$ , the number of points on $E$ is $\#E(\mathbb{F}_{1009}) = 1060 = 2^2.5.53$ , and the point $P(x, y) = (1, 237)$ which degree is $s = 530 = 2.5.53$ [1].

Let $M = 439$, and we want to apply EC-RSA-ELGamal Cryptographic System to encrypt and decrypt the message $M$

The following steps illustrate that:

- Step 1: Let $q = 23$, $r=43 \Rightarrow$ n=989, $\phi(n)$=924
- Step 2: select $1 < e < \phi(n)$, and $\gcd(e, \phi(n)) = 1$
  Suppose $e = 5 \Rightarrow$
- Step 3: $x = M^e \pmod{n}$=354 $\Rightarrow$
- Step 4: $x^3 + 71x + 602$ is a quadratic residue modulo p= 1009
- Step 5: stop and the number $M = 439$ is being mapped to the point $(x, y) = (354, 88) \in E(F_{1009})$;
  $y^2 = x^3 + ax + b \pmod{1009}$
- Step 6: select randomly;
  $a_1 = 17$, $a_2 = 432$; $1 < a_1$, $a_2 < s$ , and Compute the point $Q$ by equation (9):
  $Q(x_Q, y_Q) = A_0 + a_1 b_1 P \pmod{p}$

By using the equations (2,3,4,5) we get:

$a_1 b_1 P \pmod{p} = 281.P \pmod{1009} = (984, 175)$

$Q(x_Q, y_Q) = (354, 88) + (984, 175) = (926, 227)$

To do decryption, and coming back from the point $Q(x_Q, y_Q)$ to the number $M = 439$ , we apply the following steps:

- Step 1: Compute the point $A_0(x_0, y_0)$ as mentioned in equation (9), as following:
  $(x_Q, y_Q) + (-a_1 b_1 P) \pmod{p} = A_0(x_0, y_0) \Rightarrow$
  $(926, 227) + (-281P) = (926, 227) + (984, -175) =$
  $(926, 227) + (984, 834) = (354, 88) = A_0(x_0, y_0)$
- Step 2: Compute the message $M$ by applying:
  $x_0^d \pmod{n}$=M; $e.d = 1 \pmod{\phi(n)}$
  $x_0^d \pmod{n} = 354^{185} \pmod{989} = 439 = M$ ✓

**Note 5** We used 'JAVA Language' to compute almost all of mathematical calculations in this paper, especially which related to 'point addition and point doubling ' on elliptic curves.

### 5. Discussion and conclusion

In this paper, we showed a new approach for embedding message on an elliptic curve defined over prime fields, and this approach allows us to secure that message according to the security of the RSA Algorithm. But according to the condition; $'M < n < p'$ ,the EC-RSA-ELGamal Cryptographic System mentioned in this paper has some weakness, because the number $n$ must be more than 1024-bit according to the security of RSA algorithm, and thus the number $p$ , also, must satisfy the same, but when we working on EC over prime fields we want to select the number $p$ small as possible as we can and still keeping on the security at the same time.

In this paper, we understand that the EC-RSA-ELGamal Cryptographic System depends on two important factors for security; IFP and ECDLP, but our goal in the future is to be able to control of the condition $'M < n < p'$ , able to select the number $n$ separately from the number $p$ , and that is what we working on and hope to achieve this goal.

### Authors

**Ahmad Steef;**
PhD candidate, Department of Mathematics, Al-Baath University, Homs, Syria.

**M. N. Shamma;**
Professor in mathematics, Damascus University, Basic Sciences department, the Mechanical and Electrical Engineering, Syria, & Arab International University(AIU), Syria

**A. Alkhatib;**
Professor in mathematics, Al-Baath University, Homs, Syria.